\documentstyle[12pt,twoside,fleqn,espcrc1,psfig]{article}
\newcommand{\AmS}{{\protect\the\textfont2
  A\kern-.1667em\lower.5ex\hbox{M}\kern-.125emS}}

\title{ Pion loop fluctuations of constituent quarks and baryons}
\author{L. Ya.\ Glozman\address{Institute for Theoretical Physics, 
        University of Graz, Universit\"atsplatz 5, A-8010 Graz,  Austria},
  D. O.\ Riska\address{Department of Physics, University of Helsinki, Helsinki, Finland}}

\begin{document}
% typeset front matter
\maketitle

\begin{abstract}
The implications of the pion (meson) degrees of freedom for baryon
properties, taken at the
constituent quark and baryon levels are compared. It is shown that
there is a dramatic qualitative difference between two approaches.
\end{abstract}

\vspace{6mm}

The role of pion degrees of freedom for baryon
observables was appreciated in the strong coupling theories
of the 1950's and later formed the basis of the Skyrme model
and the chiral bag models. Consideration of the 
pion loop fluctuations on the baryon level, either
within the traditional strong coupling approach or within 
modern ChPT for baryons reveals several problems:

(i) The $\pi NN$ coupling constant is large, $g_{\pi NN}^2/4\pi \sim 14$,
which indicates that the loop expansion diverges. The expansion within 
BChPT  indicates a lack of convergence as well;

(ii) The full set of intermediate baryon resonances has to be taken into
account in the calculation, which consequently involves a large number of
parameters;

(iii) No information about the underlying quark structure is 
built into the loop integrals;

(iv) $m_N \sim \Lambda_\chi$.

\medskip
In the region of spontaneously broken chiral symmetry, where the pion 
degrees of freedom are important, the structure of the elementary
excitations of the QCD vacuum can be approximately reproduced by
absorption of the scalar interaction between bare (current) quarks in the
QCD vacuum (which is responsible for the chiral symmetry breaking)
into the mass of the quasiparticles, i.e. the constituent quarks.
Thus in the low-energy regime the proper chiral dynamics is due to
the pion-constituent quark coupling. In this case:  

(i) The $\pi QQ$ coupling constant is relatively small, 
$g_{\pi QQ}^2/4\pi \sim 0.6-0.7$;

(ii) All possible intermediate baryon resonances are taken into
account automatically by considering the $u$ and $d$ quarks
in the loop amplitudes;

(iii) The role of the underlying $SU(6)$ quark structure is 
revealed in relations between different observables;

(iv) $(m_Q /\Lambda_\chi)^2 \sim 0.1$.

\medskip

An example of the conceptual difference between the treatment of the mesonic
fluctuations of constituent quarks, on the one hand,
 and (directly) of baryons, on the other hand, is in order.
In the latter the mesonic loops imply an infinite shift of
the baryon mass in the chiral limit, which is balanced by the
phenomenologically determined counter-terms. Consideration of the
mesonic degrees of freedom in such approaches yields information
on the corrections from the finite masses of current quarks, but
not on the origin of the nucleon mass nor on the octet-decuplet
splitting in the chiral limit. In the chiral constituent quark model
the role of the meson degrees of freedom is broader. On the one hand
meson fluctuations contribute to the self-energy of the constituent
quarks (loops at the quark level), but on the other hand they imply a strong flavor-dependent
interaction between the constituent quarks (meson exchange interaction
between different quarks), which yields an octet-decuplet
splitting in the chiral limit \cite{G1}. 

\medskip

The pion-exchange interaction contains both the ultraviolet (short-range) 
part,
which is independent of the pion mass, and the infrared 
(Yukawa) part. The latter one is important for the long-range
nuclear force, but it does not produce any significant effect
in the baryon because of its small matter radius. The short-range
part of the pion-exchange interaction produces a flavor-spin
dependent force between quarks and has a range $\Lambda_\chi^{-1}$.
While the infrared (Yukawa) part of the interaction vanishes
in the chiral limit, the ultraviolet part - does not. This
means that in some sense the short-range part of the pion-exchange
interaction is "more fundamental" than its Yukawa part. Note
that this short-range interaction stems from the $\gamma_5$
structure of the pion-quark vertex (i.e. exclusively from the
quantum numbers of the pion) and hence is demanded by the
Lorentz invariance. This short-range part of the interaction,
combined with the $SU(6)$ structure of the zero order baryon
wave functions (which is demanded by the large $N_c$ limit in QCD),
provides a basis for the explanation of the low-lying baryon spectrum
\cite{SPECTRUM}. It should be noted that this short-range interaction
cannot be obtained by considering the pion loops at the baryon
level with the ground $N$ and $\Delta$ states as the intermediate
states within the loop \cite{G2}. At the baryon level one needs to
consider the whole infinite tower of the radially excited states
within the loop in order to meet this short-range meson exchange
interaction. This implies that the meson-baryon or quark
models which employ only the subspace of the $\pi N$ and $\pi \Delta$
states are very incomplete and therefore do not take
into account the most important
short-range effects of the pion (meson) degrees of freedom for
baryon structure.

\medskip
As another example consider the different role 
of the pion cloud for
magnetic moments in both approaches. 
In hadronic models the pion loops is the
only source for the nucleon anomalous magnetic moment. In contrast,
within the naive quark model the nucleon magnetic moments are
exclusively due to the intrinsic $SU(6)$ quark structure of
the baryon. The small pion-quark coupling constant together
with the additional small parameter $(m_Q /\Lambda_\chi)^2 \sim 0.1$
guarantee that the loop corrections to the naive quark model
predictions (i.e. loops at the constituent quark level)
are not large \cite{GR}. For the absolute values of the
proton and neutron anomalous magnetic moments one finds contributions
at the level of 5-10\%. What is more important, the ratio $\mu_n/\mu_p$,
comes out to be only 2\% above the empirical ratio 0.68, while
the naive quark model prediction, $2/3$, is 2\% below.

\medskip
The pion loop corrections supply only a tiny contribution to the
neutron Dirac charge radius \cite{GR}. They therefore do not 
perturb the satisfactory description of the negative mean square radius of the
neutron, which is implied by the empirical value of the neutron
magnetic moment (Foldy term).

\medskip
The kaon loop fluctuations of the valence $u$ and $d$
constituent quarks  induce only a very small
strangeness magnetic moment of the proton \cite{HRG}, which is
in the range -0.01 -- -0.05, and is well consistent with
the most recent data from the SAMPLE experiment, reported at
this conference. This is again in contrast with the large negative
strange magnetic moment that is implied by the kaon loop
fluctuations considered at the baryon level.

\end{document}